\documentclass[12pt,preprint,epsfig]{aastex}
\begin{document}

\title{Analysis of two quintessence models \\
with SN Ia data}

\author{M. Pavlov\altaffilmark{1}}
\affil{Osservatorio Astronomico di Capodimonte, via Moiariello
16, I-80131 Napoli, Italy}

\author{C. Rubano}
\affil{Dipartimento di Scienze Fisiche - Univ. Federico II di Napoli,
INFN Sez. di Napoli, Complesso Universitario di Monte S. Angelo, Via
Cintia, ed. G, I-80126 Napoli, Italy}
\email{rubano@na.infn.it}

\author{M. Sazhin}
\affil{Sternberg Astronomical Institute,
Universitesty pr. 13, Moscow 119899, Russia}

\and

\author{P. Scudellaro}
\affil{Dipartimento di Scienze Fisiche - Univ. Federico II di Napoli,
INFN Sez. di Napoli, Complesso Universitario di Monte S. Angelo, Via
Cintia, ed. G, I-80126 Napoli, Italy}

\altaffiltext{1}{Permanent Address: Sternberg Astronomical Institute,
Universitesty pr. 13, Moscow 119899, Russia}

\begin{abstract}
The supernovae Ia data are used to analyze two general exact solutions for
quintessence models. The best fit values for $\Omega _{m0}$ are smaller than
in the $\Lambda $-term model, but still acceptable. With present-day data,
it is not possible to discriminate among the various situations.
\end{abstract}

\keywords{cosmology: quintessence, supernovae: type Ia supernovae}

\section{Introduction}

Recently, astronomers discovered an accelerated expansion of our Universe.
It is well known that all known types of matter generate attraction, which
leads to a decelerated expansion of the Universe. That discovery then
reveals a new type of matter, which is now called {\em quintessence} or,
sometimes, {\em dark energy} \citep{ost95,tur97,chi97,cal98,zla99,pmw99}.

The discovery of the presence of dark energy became possible when
astronomers recognized that SN Ia can be the long expected {\em standard
candle} for cosmological investigations. Two main features provide the use
of SN Ia as a standard candle \citep{fil00}:

{\em i}) They are exceedingly luminous, comparable with luminosity of a
whole galaxy; they can, thus, be detected and observed with high S/N ratio
even at cosmological distances.

{\em ii}) ``Normal'' SNe Ia have small variations among their peak
absolute magnitudes (around 0.3).

The accelerated expansion of our Universe was discovered as a result of
two projects: the High-z SN Search \citep{sch98,rie98} and the Supernova
Cosmology Project \citep{per99}.

In fact, a new type of matter was predicted many years ago by A. Einstein,
who included a $\Lambda$-term into his considerations \citep{ein17}. At
the beginning of the past century, the $\Lambda$-term was just a new
fundamental constant, and only much later it was really considered as a
formidable challenge by both observational and theoretical cosmologists
\citep{wein89,zel90,zel92,cpt92,carr00,star00,rub00}. Moreover, during the
last 20 years cosmologists understood that this constant can be replaced
with a scalar field, which induces the repulsive gravitational force
dynamically. Accordingly, several models were proposed
\citep{rat88,pee88,wet95,wet98,cop98,fer98,lid99,ste99,bra99,sah00,bin00,rub01},
in order to explain the observed present acceleration of our Universe.

Two of these models continued to be developed after the discovery of
acceleration, and were also roughly elaborated and adapted for present-day
data \citep{rub01}. Here, we again use these models to such an end, but in
a much more refined way: the goal is now to fit the observed data of
apparent magnitude and redshift of the supernovae Ia, and test the models
themselves.

\section{Model description}

As said above, in this paper we discuss two models for quintessence, both
based on a scalar field with a special type of potential. The field is
minimally coupled with pressureless matter, and the total density
parameter $\Omega$ of the Universe is fixed to be 1. A detailed discussion
of the consequences of assuming such models in cosmology is given in
\citep{rub01}, so that we limit ourselves here only to a short summary of
the results we need for our purpose.

The main attractive feature of these models is that they allow a general
exact solution of the field equations, obtained through a suitable
transformation of variables. Anyway, independently of the fact that this is
an exact solution, we also find that this solution reflects many properties
of the real Universe correctly.

The first model considers a potential of the form
\begin{equation}
V(\varphi) =B^2 e^{-\sigma \varphi},  \label{eq1}
\end{equation}
where $B^2$ is a generic positive constant and $\sigma^2$ is some fixed
combination of universal constants
\begin{equation}
\sigma^2 = {\frac{\displaystyle 12 \pi G}{\displaystyle c^2}}.  \label{eq2}
\end{equation}

Actually, this kind of potential has already been widely discussed in the
literature, but without any particular assumption on the value of $\sigma$
\citep{rat88,pee88,wet95,wet98,cop98,fer98,bra99,sah00,bin00,fab00}. We
stress that it is the particular choice of this constant given above that
allows the exact integration of the field equations (see also
\citep{bar87,bur88}).

The general solution of the cosmological equations (both for the metric
and the scalar field) has five free parameters (including $B^2$)
\citep{rub01}. We fix two of them and keep three as free. But one of these
three parameters just determines the present value of the scale factor of
our Universe, which, in a spatially flat geometry, is not observable. It
is not included into statistical analysis and does not affect the degrees
of freedom of our analysis.

We list below only the three cosmological functions which we need in our
analysis (the other ones can be found in \citep{rub01}, of course)
\begin{eqnarray}
(1 +z)^3 ={\frac{\displaystyle \tau^2_0(1+ \tau^2_0)}{\displaystyle \tau^2
(1 + \tau^2)}},  \label{eq3} \\
\phantom{blanck space}  \nonumber \\
H={\frac{\displaystyle 2(1 + 2\tau^2)}{\displaystyle 3t_s\tau (1 +\tau^2)}},
\label{eq4} \\
\phantom{blanck space}  \nonumber \\
\Omega_m = {\frac{\displaystyle 1 + \tau^2}{\displaystyle (1 + 2\tau^2)^2}}.
\label{eq5}
\end{eqnarray}
They are the redshift of the epoch, the Hubble parameter, and the
$\Omega_m$ parameter of pressureless matter, and are expressed in terms of
the dimensionless time $\tau \equiv t/t_s$. The free parameters are then
the time scale $t_s$ and the present value of the dimensionless time
$\tau_0$. Le us remark that $t_s$ is of the same order of magnitude (but
not necessary equal to) as the age of the Universe.

As to the second model, it considers a potential of the form
\begin{equation}
V(\varphi) = A^2 e^{\sigma \varphi} +B^2 e^{-\sigma \varphi},  \label{eq6}
\end{equation}
with $\sigma^2 =12 G \pi / c^2$ as before, and $A^2$ and $B^2$ free
parameters.

We have, now, one additional free parameter; therefore, according to the
same considerations as above, we have to deal with three of them.

The equations which describe the Hubble parameter, the redshift, and the
density parameter in the second model are
\begin{eqnarray}
(1+z)^3 = {\frac{\displaystyle \lambda^2 \sinh^2 \tau_0 \sin^2 \tau_0}{%
\displaystyle \lambda^2 \sinh^2 \tau -\sin^2 \tau}},  \label{eq7} \\
\phantom{blanck space}  \nonumber \\
H(\tau)={\frac{\displaystyle \omega (\sin(2 \tau) -\lambda^2 \sinh (2\tau))}{%
\displaystyle 3(\sin^2 \tau -\lambda^2 \sinh^2 \tau)}},  \label{eq8} \\
\phantom{blanck space}  \nonumber \\
\Omega_m(\tau) = {\frac{\displaystyle 2(\lambda^2-1)(\cos(2\tau) +\lambda^2
\cosh(2\tau) -1 - \lambda^2)}{\displaystyle (\sin(2\tau) -\lambda^2
\sinh(2\tau))^2}}.  \label{eq9}
\end{eqnarray}
The dimensionless time, in this case, is $\tau=\omega t$. Following
\citet{rub01}, we use here $\omega$ instead of $t_s$, because of the fact
that it is directly connected with the parameters in the potential of the
scalar field, and has the meaning of a mass factor in theoretical
considerations. So, the free parameters are $\omega$, $\lambda$, and
$\tau_0$.

In the analysis of the supernovae data, we use the bolometric distance. As
explained better below, it can be expressed in terms of the ``Hubble
free'' luminosity distance and of a parameter $m_0$, connected with the
absolute magnitude and the Hubble parameter. The parameters of the first
model ($\tau_0$ and $t_s$) can be recasted into $\tau_0$ and $m_0$. The
parameters of the second model ($\lambda$, $\tau_0$, and $\omega$) can be
recasted into $\lambda$, $\tau_0$, and $m_0$. Once the best fit is made,
it is easy to compute the relevant physical quantities $H_0$ and
$\Omega_{m0}$. In all the considerations below, $H_0$ turns out to have
the same value as in \citep{per99}. So, we concentrate on $\Omega_{m0}$.

\section{SNe Ia Data}

The published data of the supernovae consist of 60 SNe Ia \citep{per99}.
The data analysis and the determination of cosmological parameters can be
considered in two steps. The first one is the measurement of the Hubble
parameter for close supernovae (Calan - Tololo survey) \citep{ham96}, to
be compared with the absolute magnitude $M$ of a supernova SN Ia. The
second step is the comparison of the high redshift supernovae with the
theoretical prediction of bolometric distance:
\begin{equation}
m=5\log (D_b) + m_0;  \label{eq10}
\end{equation}
here, $D_b$ is the "Hubble free" bolometric distance
\begin{equation}
D_b = H_0(1 + z)\int^z_0\frac{dz^{\prime}}{H(z^{\prime})},  \label{eq11}
\end{equation}
and $m_0$ is a parameter connected to the absolute magnitude and the
Hubble parameter.

In data presented in \citep{per99} there are several values for corrected
apparent magnitude. Authors consider $m^{peak}_B$ and stretch luminosity
corrected effective $B$-band magnitude $m^{eff}_B$. For the analysis of
cosmological parameters only $m^{eff}_B$ is used, together with its errors
$\sigma_{m_B^{eff}}$.

There are several methods for SN Ia data analysis. Two of them are used in
\citep{rie98}. The first one is the Multicolor Light Curve Shape (MLCS)
method and the second one is a template fitting method. In \citep{per99}
another method is used.

The data of both groups have the statistical errors approximately as
$\sigma_m \sim 0.25$.

We follow the authors of paper \citep{per99} to analyze the models
described in \citep{rub01}. First of all, as a check of the procedure, we
apply the flat cosmological model with a $\Lambda$-term to fit the data.
The standard $chi^2$ algorithm of data analysis reveals a good agreement
of our analysis with published statistical values \citep{per99}. We use
the complete set of data of 60 SNe Ia. It results $\chi^2=1.75$ per degree
of freedom, not significantly different from $\chi^2=1.76$ found in
\citep{per99}. The same is for the $\Omega_\Lambda$ parameter. Since 4
points in the data are outliers, we can proceed with analysis and exclude
these data from our considerations. The total number of SN Ia data then
drops to 56. The $\chi^2$ per degree of freedom in this case becomes 1.16,
which is in good agreement with previously published results \citep{per99}
and is within $1\sigma$ level.

\section{Data analysis and fitting}

In our analysis we use the standard $\chi^2$ method. The analysis is done
minimizing the value of weitghed $\chi^2$:
\begin{equation}
\chi^2 = \sum w_i (m_i - m^{model}_i)^2,
\end{equation}
where $w_i$ is the weight of the $i$-th SN Ia, $m_i$ is its $B$-band
effective apparent magnitude, and $m^{model}_i$ is its magnitude as
predicted with the models introduced before and thoroughly discussed in
\citep{rub01}.

\subsection{The first model}

In the first model, it is possible to eliminate $\tau$ from Eqs.
(\ref{eq3}) and (\ref{eq4}), and to obtain an analytical expression for
$H(z)$. Thus, it is possible to compute $m$ from Eqs. (\ref{eq10}) and
(\ref{eq11}), and $\chi^2$ as a function of $\tau$ and $m_0$.

Firstly, we use 60 SN Ia data and get the $\chi^2$ minimum at $m_0=24.01$,
$\tau_0=1.04$, with $\chi^2 = 1.77$ per degree of freedom. As it is
unsatisfactory, we reject data which are out of the 3$\sigma$ level, as
done in \citep{per99}.

After data rejection, the $\chi^2$ minimum drops down to $\chi^2=1.195$
per degree of freedom. It is definitely within one sigma level of the
expected value of $\chi^2$. The minimum now has other values than
$m_0=23.985$ and $\tau_0=1.268$.

If we accept the value of this minimum, we obtain, from Eqs. (4) and (5),
$H_{0}=70kms^{-1}Mpc^{-1}$, $\Omega _{m0}=0.15$.

The situation is illustrated in Figs. 1 -- 3.

\subsection{The second model}

The second model has only been tested and fitted with 56 data of SNe Ia.
The number of parameters in this case is equal to three. The true minimum
of the $\chi ^{2}$ is at $m_{0}=23.98$, $\tau _{0}=0.8$, and $\lambda
=1.182$. We find a value of $\chi ^{2}=1.1906$, which is definitely within
one sigma level of expected value. After such value we obtain from Eqs.
(\ref{eq8}) and (9) that $\Omega _{m0}=0.17$.

The $\chi ^{2}$ value is a function of three arbitrary values: $m_{0}$,
$\tau _{0}$, and $\lambda $. Therefore, the $\chi ^{2}$ as a function of
all parameters is impossible to plot, but we can nonetheless plot several
slits.

The situation is illustrated in Figs. 4 -- 8.

\section{Conclusions}

In a quintessential universe we have analyzed the same data as in
\citep{per99}, where it is present only a cosmological constant, and found
good values for $\chi^{2}$ in both cases. The values of $\Omega_{m0}$
found are rather different from the one found in \citep{per99}, but it is
impossible to say if this is due to differences in the models or to
influence of the measurement errors on the final values.

In fact, in both models we have degeneracy in the parameters, particularly
large in $\lambda$ (II model). This makes impossible to give significant
confidence limits for the values of $\Omega _{m0}$, which we found. Only
very rough estimates can be given. Our main results are summarized in the
following tables.

\bigskip

\begin{center}
Model I

\medskip

\begin{tabular}{|c|c|c|c|c|c|}
\hline
$\chi ^{2}$ & $m_{0}$ & $\tau _{0}$ & $\Omega _{m0}$ & $\tau _{0}$ range & $%
\Omega _{m0}$ range \\ \hline
1.195 & 23.985 & 1.268 & 0.15 & 0.82 $\div $ 1.40 & 0.12 $\div $ 0.30 \\
\hline
\end{tabular}

\bigskip

Model II\medskip

\begin{tabular}{|c|c|c|c|c|c|}
\hline
$\chi ^{2}$ & $m_{0}$ & $\tau _{0}$ & $\lambda $ & $\Omega _{m0}$ & $\Omega
_{m0}$ range \\ \hline
1.19 & 23.98 & 0.8 & 1.182 & 0.17 & 0.14 $\div $ 0.22 \\ \hline
\end{tabular}
\end{center}

As final remarks, we want to observe that our results are in a very good
agreement with the one found in \citep{bah00} in a completely independent
way, and that the high degeneracy we get for the model parameters seems to
support the opinion of those who claim that it is very difficult to
discriminate among theories on the basis of observational data only
\citep{mao00,bar00}.

\acknowledgments

M.V. Sazhin acknowledges the INFN (Naples Section) and Osservatorio
Astronomico di Capodimonte for financial support and hospitality during
his visit in Napoli. M.V. Pavlov acknowledges the support of Osservatorio
Astronomico di Capodimonte. C. Rubano and P. Scudellaro are in part
supported by MURST Prin2000 SINTESI.

\clearpage

\begin{figure}
\plotone{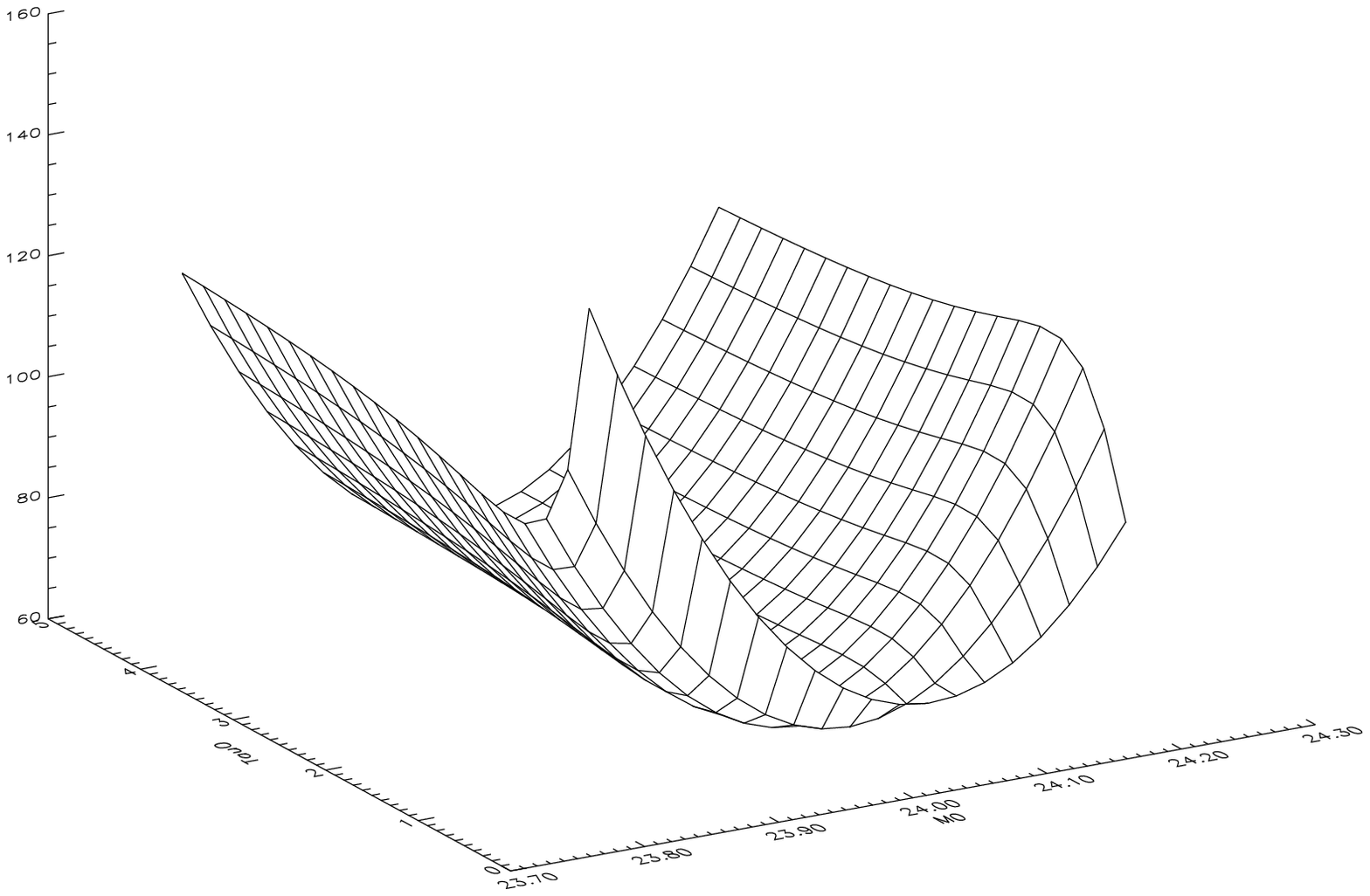}
\caption{The surface of $\protect\chi ^{2}$ as a function of two variables,
the $\protect\tau _{0}$ parameter and $m_{0}$ parameter. This function has
a very definite and sharp minimum.
\label{fig1}}
\end{figure}

\begin{figure}
\plotone{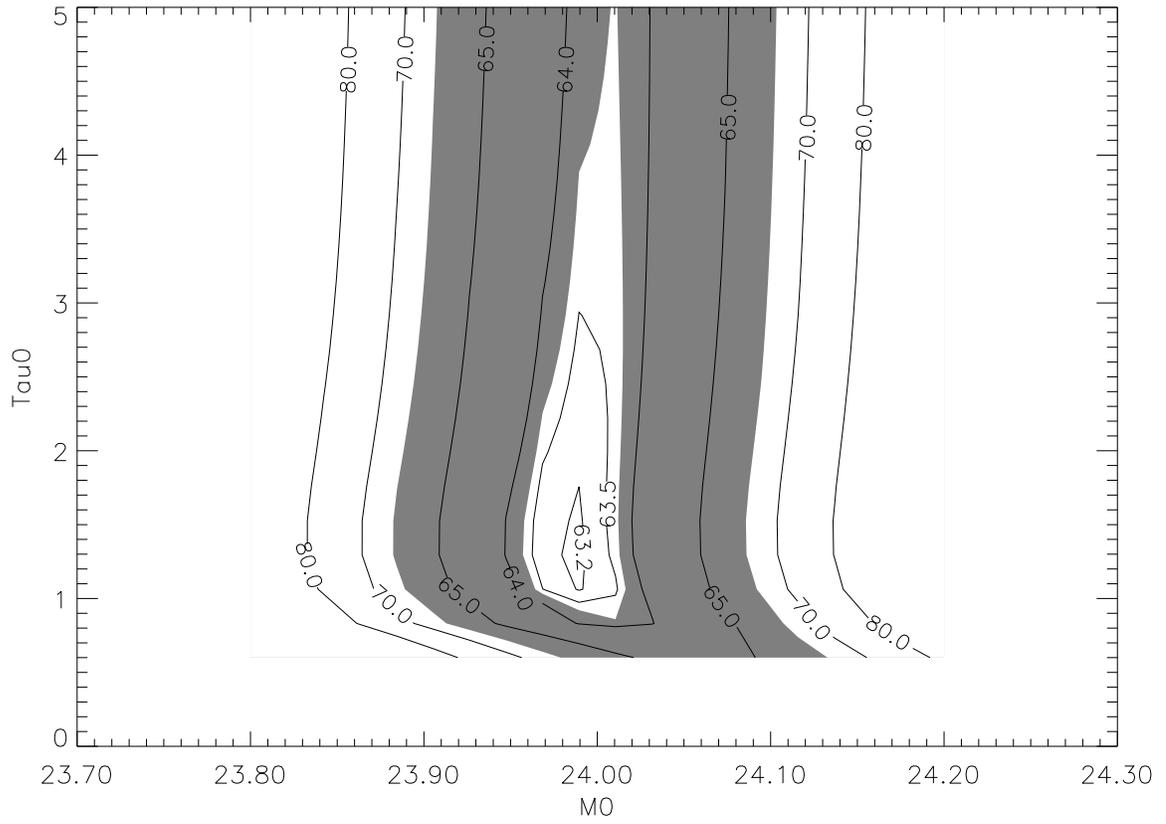}
\caption{Countour plot of the levels of $\protect\chi ^{2}$ is shown. The
one sigma level is the small white region on the graph. The two sigma level
is the shadowed region. It reveals a large degeneracy of the $\protect\chi %
^{2}$ function with respect to the parameter $\protect\tau _{0}$.
\label{fig2}}
\end{figure}

\begin{figure}
\plotone{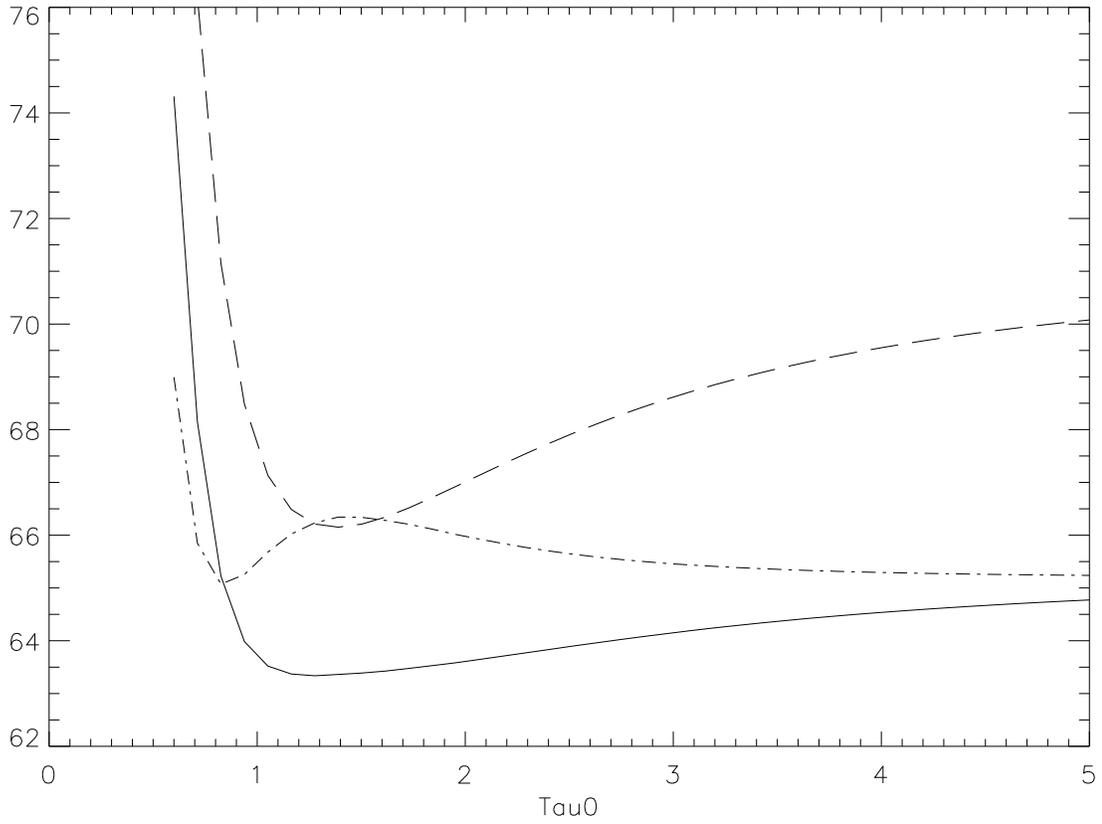}
\caption{The solid line corresponds to the slit of 2D $\protect\chi ^{2}$
taken at the point of true minimum $m_{0}=23.985$. The dot-and-dashes curves
corresponds to the slit taken at $m_{0}=24.035$, and the dashes curve to the
slit taken at $m_{0}=23.935$. It results that the minimum in $\protect\tau %
_{0}$ is strongly dependent on the $m_{0}$ minimum.
\label{fig3}}
\end{figure}

\begin{figure}
\plotone{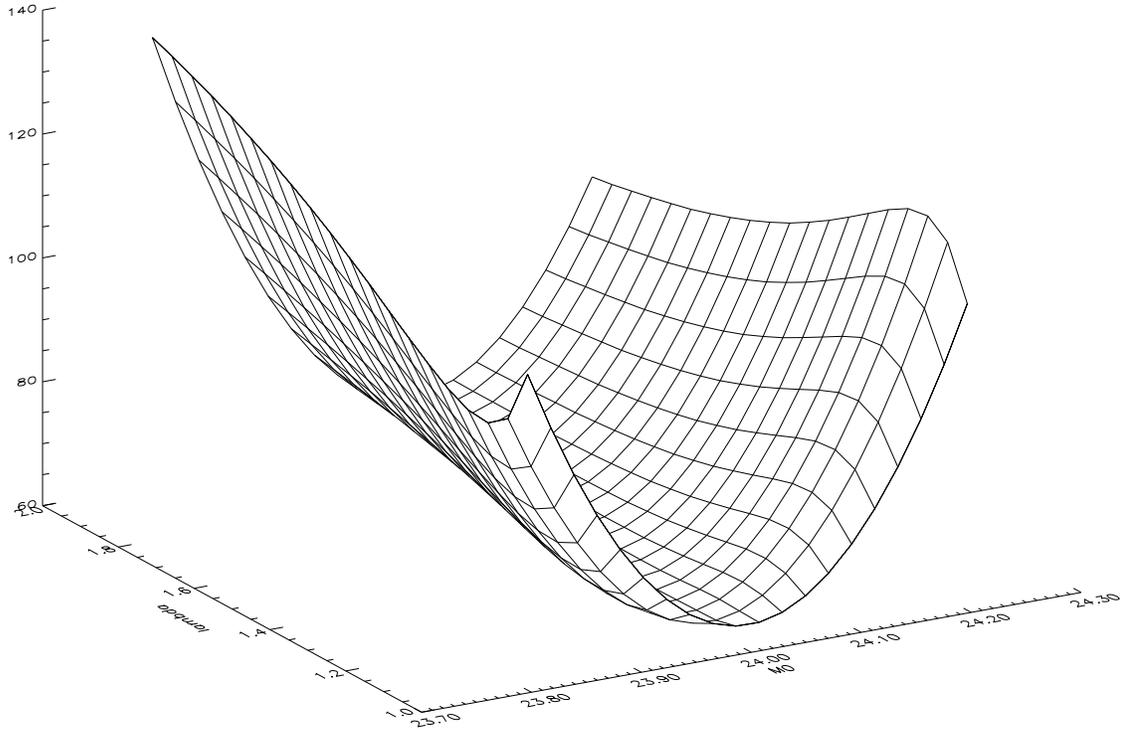}
\caption{The surface of $\protect\chi ^{2}$ as a function of two parameters:
$m_{0}$ and $\protect\lambda $; here, we fix $\protect\tau _{0}=0.8$. One
can see the profile of this surface and minimum.
\label{fig4}}
\end{figure}

\begin{figure}
\plotone{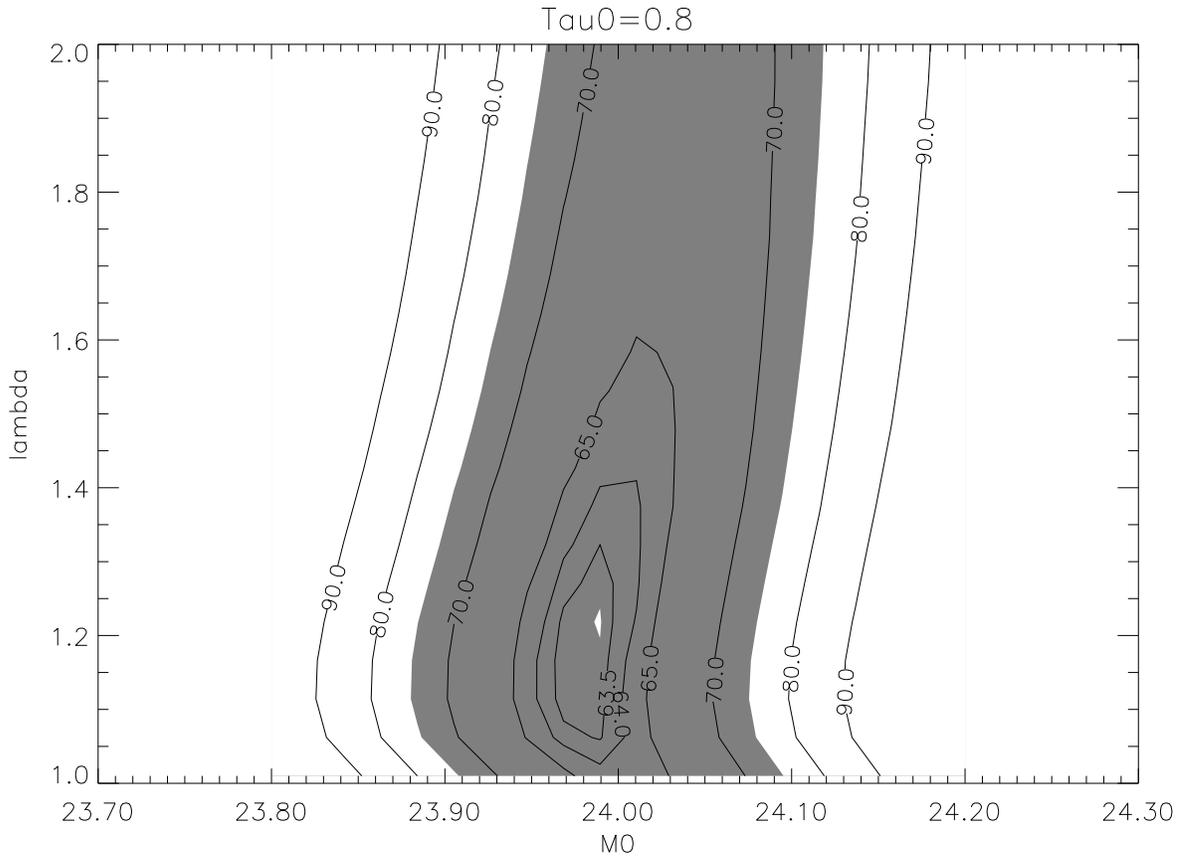}
\caption{Contour plot of the Fig. 4 surface. The 1$\protect\sigma $ area is
the small white region at $m_{0}=23.985$ and at $\protect\tau _{0}=1.268$.
The shadowed region is the $2\protect\sigma $ area, revealing that there
is parameter degeneracy.
\label{fig5}}
\end{figure}

\begin{figure}
\plotone{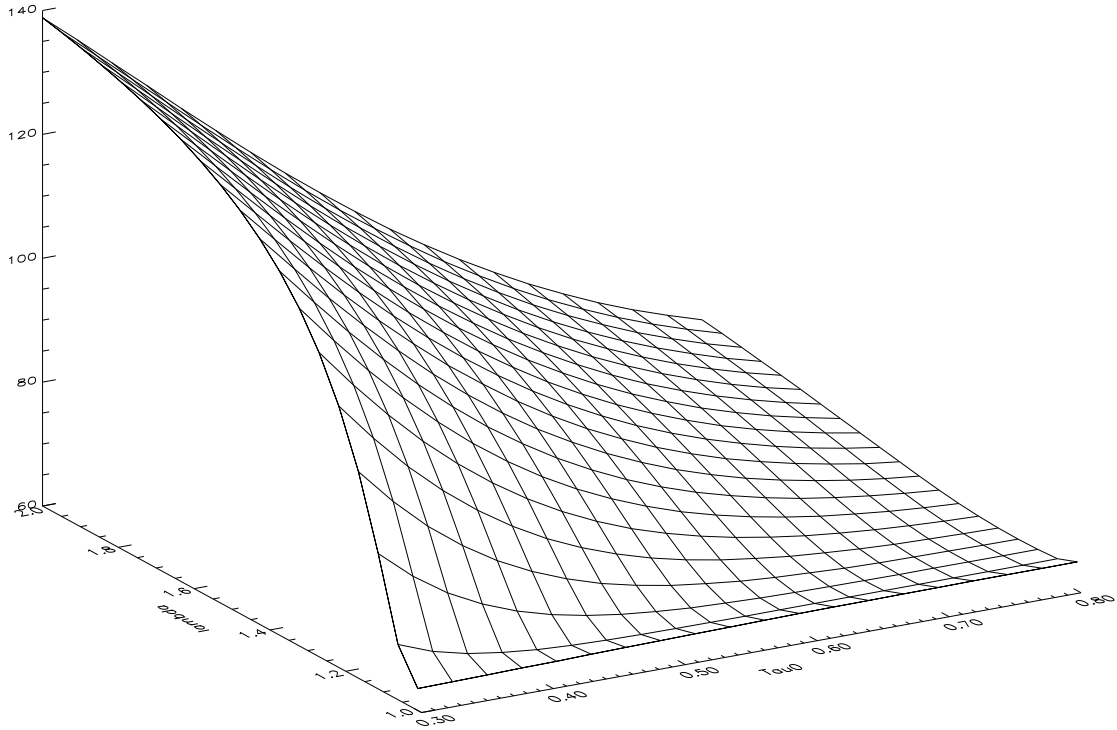}
\caption{We plot $\protect\chi ^{2}$ as a function of other two parameters, $
\protect\tau _{0}$ and $\protect\lambda $, fixing $m_{0}=23.985$.
\label{fig6}}
\end{figure}

\begin{figure}
\plotone{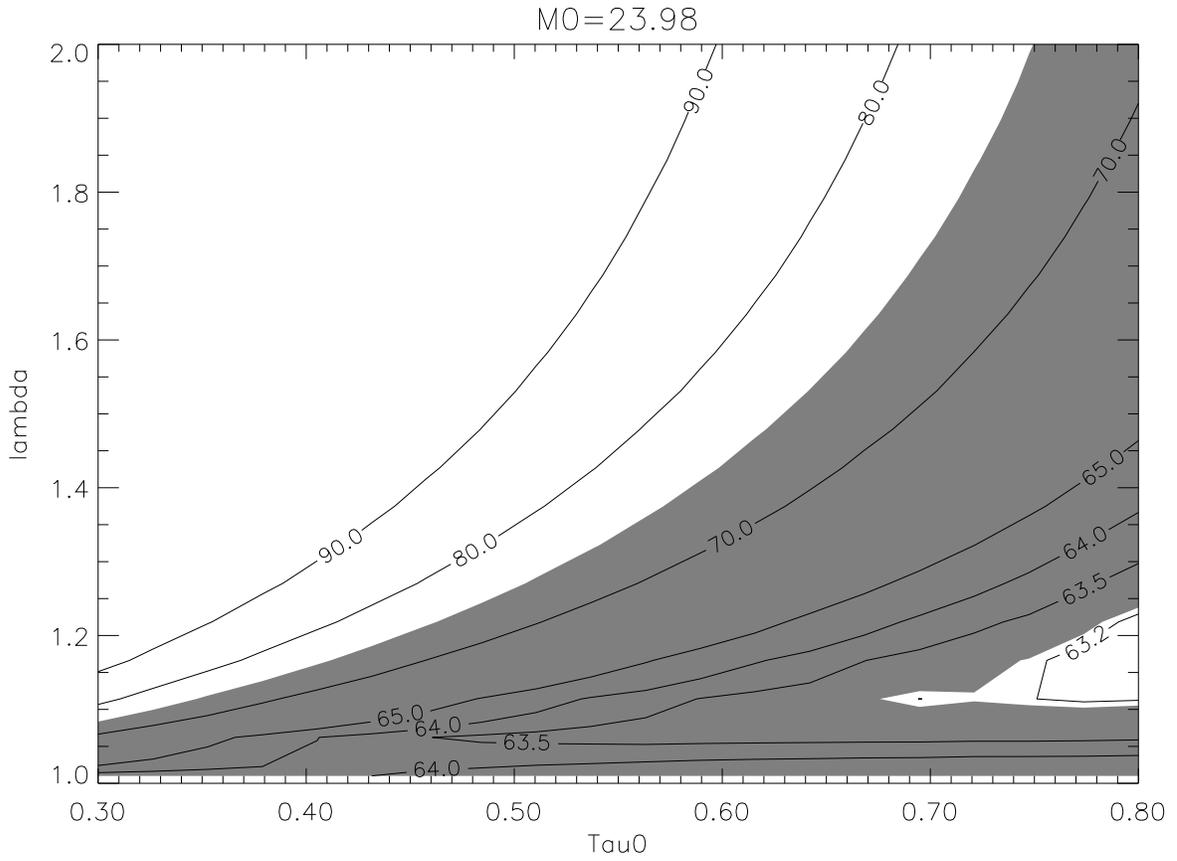}
\caption{Countour plot of Fig. 6 surface. The white region in the right
corner below is the $1\protect\sigma $ level of parameters $\protect\tau
_{0} $ and $\protect\lambda$. The shadowed region is the $2\protect\sigma $
level of these parameters. One can see again a large degeneration for
them.
\label{fig7}}
\end{figure}

\begin{figure}
\plotone{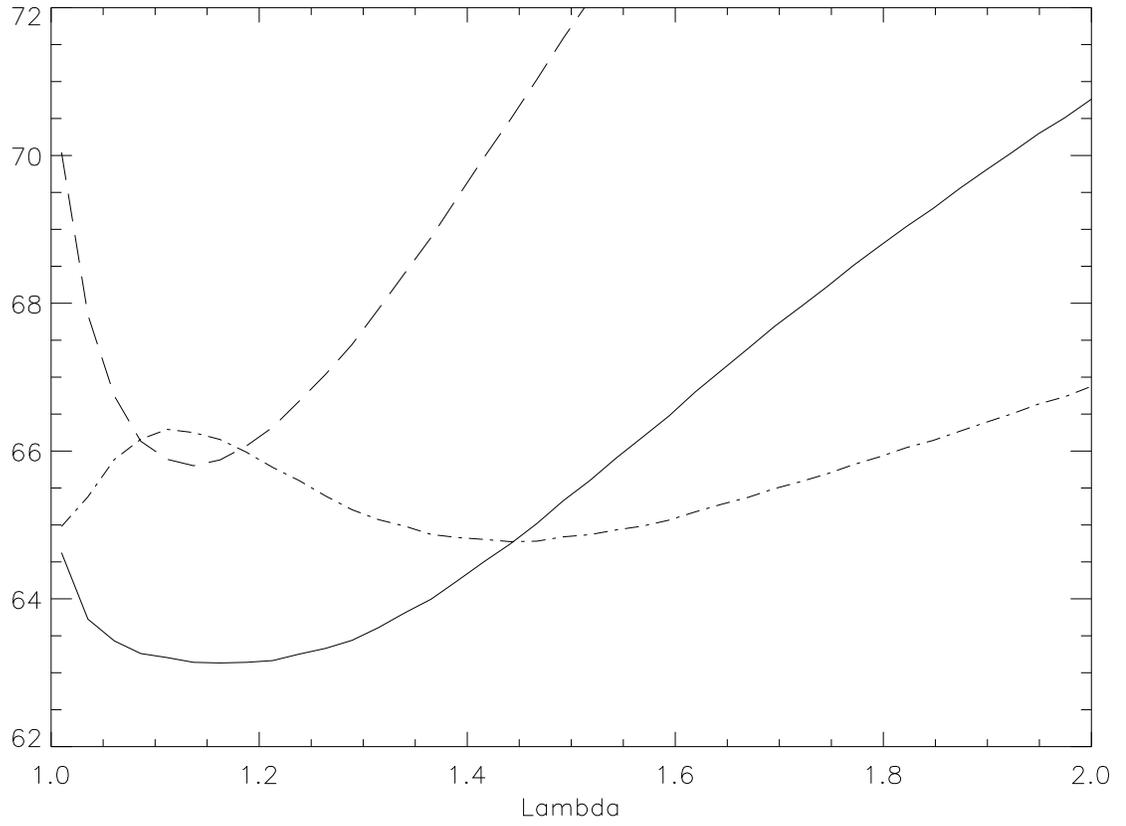}
\caption{Three different slits at fixed $m_{0}=23.985$ and $\protect\tau
_{0}=0.8$, showing degeneracy in $\protect\lambda $.
\label{fig8}}
\end{figure}

\end{document}